\documentclass[12pt,letterpaper]{article}
\usepackage{graphicx,latexsym}
\usepackage[utf8]{inputenc}
\usepackage{hyperref}
\usepackage{fullpage, setspace} 
\usepackage{enumerate}
\title{Site Reliability Engineering(SRE) and Observations on SRE Process to make tasks easier}
\begin{document}
        \date{}  
	\maketitle 
	\begin{center}
        Balaram Puli \\
        Independent Researcher \\
        balaram.puli@ieee.org \\
	\end{center}	
	\begin{center}
          \vspace{-1em}
          July 13, 2022
        \end{center}
	\section*{Introduction}
	 Site Reliability Engineering (SRE) Implements DevOps practice, the goal of SRE is to accelerate product development teams and keep services running in reliable and continuous way in the production environments whether it could be on-premise or cloud environment. SRE helps to increase sharing and collaboration between the Development and Operations team and helps to resolve the issues very quickly. In the production environment issues are Normal and more troubleshooitng is needed. SRE practice  is more profitable to focus on speeding recovery of application than preventing accidents/issues. \newline
	 
	 SRE will focus on small and gradual changes and This SRE strategy, coupled with automatic testing of smaller changes and reliable rollback of bad changes when deployments get failed, leads to approaches to Change Management like CI/CD(continuous integration and continuous Deployment) on on-premise and cloud environments.\newline
	 
	 The main SRE Foundations are SLIs, SLOs, and SLAs Monitoring, Alerting.\newline
	 	 
	 SLIs(Service Level Indicators) are metrics over time and specific to a user journey such as request/response, data processing, which shows how well the service is performing/functioning. SLIs is the ratio between two numbers which are the good and the total,
	 \begin{enumerate}[a)]
	 	\item Success Rate = No. of successful HTTP request / total HTTP requests
	 	\item Throughput Rate = No. of consumed jobs in a queue / total number of jobs in a queue
	 \end{enumerate}
	SLI is divided into specification and implementation, for example
	\begin{enumerate}[a)]
		\item Specification: ration of requests loaded in $<$ 100 ms
		\item Implementation is a way to measure, for example based on a) server logs  and b) client code instrumentation.
	\end{enumerate}

	 SLI ranges from 0\% to 100\%, where 0\% means nothing works, and 100\% means nothing is broken on the environemnt or Application.\newline
	 SLO (Service Level Objective) is a target percentage based on SLIs and can be a single target value or a range of values for example, SLI $<$= SLO or (lower bound $<$= SLI $<$= upper bound) = SLO \newline
	 
	Error budgets are a tool for balancing reliability with other engineering work, and a great way to decide which projects will have the most impact. An Error budget is 100\% minus the SLO \newline
	
	Monitoring allows you to gain visibility into a system, which is a core requirement for checking a service health and diagnosing the service when things are going in a wrong or out of control.\newline
	
	from an SRE perspective,
	\begin{itemize}
		\item Alert on conditions that requires attention to take immediate action.
		\item Investigate and diagnose issues of application, system, environment.
		\item Display information about the system visually to understand more granular level.
		\item Gain insight into the system health and resource usage for long-term planning.
		\item Compare the behavior of the system before and after a change or enhancement of the application, or between two control groups. \newline
	\end{itemize}

	Monitoring features that might be relevant,
		\begin{itemize}
		\item Speed of data retrieval and freshness of the data and analysis.
		\item Data retention  and calculations.
		\item Interfaces like graphs, tables, charts can be High level or low level for deeper analysis.
		\item Alerts: multiple categories, notifications flow, suppress functionality.
	\end{itemize}

	Alerting helps ensure alerts are triggered for a significant event or incident, an event or incident that consumes a large fraction of the error budget. Alerting should be configured to notify an on-caller only when there are actionable, specific threats to the error budget along with other application issues.\newline
	Alerting considerations:
	\begin{itemize}
		\item Precision: The proportion of events detected that were significant.
		\item Recall: The proportion of significant events detected.
		\item Detection time: How long it takes to send notification in various conditions. Long detection time  negatively impacts the error budget.
		\item Reset time: How long alerts fire after an issue is resolved.
	\end{itemize}

	The proposed approach is that SRE(Site Reliability Engineering) practices require a significant amount of time and skilled SRE people to implement best process and A lot of tools are involved in day-to-day SRE work to mitigate the production incidents/issues. SRE processes are one of a key mechanism to the success of a tech company.\newline

	\section*{Problem Statement:}
	Troubleshooting is a critical skill for anyone who operates on distributed computing systems especially SREs, but it’s often viewed as an niche skill that some people have and others don’t. One reason for this assumption is that, for those who troubleshoot very often, it’s an ingrained process and explaining how to troubleshoot and it's difficulties, we believe that the troubleshooting is both learnable and teachable.\newline
	
	Every problem starts with a problem report which might be an automated alert or one of your colleagues saying that “The system is slow, running with HIGH CPU and Memory.” An effective report should tell you the expected behavior and the actual behavior of the application, also gives an information on  how to reproduce the behavior of the application  under certain memrory,CPU, Load. Ideally, the problem reports or Root Cause Analysis(RCA) should have a consistent form and be stored in a searchable location for customers, such as a bug tracking system. Here, our teams often have customized forms or small web apps that ask for information that’s relevant to diagnosing the particular systems they support, which then automatically generate and route a bug. This may also be a good point at which to provide tools for problem reporters to try self diagnosing or self-repairing common issues on their own.\newline
	
	It is common practice at Google to open a bug for every issue irrespective of big or small issue, even those received via email or instant messaging. Doing so creates a log of investigation and remediation activities that can be referenced in the future. Many teams discourage reporting problems directly to a person for several reasons: this practice introduces an additional step of transcribing the report into a bug, produces lower-quality reports that are not visible to other members of the team, and tends to concentrate the problem solving load on a handful of team members that the reporters happen to know, rather than the person currently on duty.\newline	
	
	\section*{Objectives:}
	We will extend previous work on SRE by overcoming 4 of its current limitations,\newline
	\begin{itemize}
	\item Firstly, create granular level metrics on applications with respect to CPU, disk, RAM, throughput , latency, transactions per second and so on and so forth.
	\item Secondly, closely monitoring system has to be implemented for every on-premises or cloud based application. 
	\item Thirdly, Deep logging must be enabled for every application and it should be available for troubleshooting and triaging the issue.
	\item Finally, perform the postmortem on the issue by collecting all the kinds of application and dependent subcomponent logs, traces, heap dumps and thread dumps and so on and so forth. 
	\end{itemize}
	
	\section*{Preliminary Literature Review}
	Site Reliability Engineering (SRE) is an area of the utmost prominence for software based systems active in safety-critical applications such as computer relaying of power transmission lines or online web applications. Site Reliability is the probability of failure-free software operation for a specified period of time in a specified environment whether it could be on-premise or cloud envronment. Site Reliability is an important factor affecting system reliability and monitoring the production applications very closely and work together with DevOps teams. 	
	The high density of software is the major causative factor of software reliability snags. Site reliability is a measuring technique for flaws that causes software failures in which software behavior is different from the specified/original behavior in defined environment with fixed time. Various approaches can be used to improve the reliability of software, however, it is hard to balance development time and budget with Site reliability. A critical review of the software and their reliability is most imperative and the prerequisite of the software era. This paper inculcates the consistency of the software on reveal basis of literature survey and production system application monitoring, alerting, SLIs, SLAs, SLOs.\newline
	
	Ensuring that the information of hardware and software systems reliability is a prerequisite for Site Reliability Engineering. SRE is possible only with the software components adaptability, 
	that is, the ability to find and eliminate software and hardware errors. By the elimination criterion, errors or failures can be divided into recoverable and non-recoverable errors. Fatal errors make it possible to reduce the failure rate and thereby increase reliability of applications. The recoverable errors accumulation leads to the accumulated reliability concept. Accumulated reliability is higher than the information and software systems initial 
	reliability. The improving reliability proposed method is applicable to any adaptive systems in which it 
	is possible to identify and eliminate errors or malfunctions in the production applications. This approach is applicable to intelligent systems and technologies in which the mechanism for eliminating failures or errors is fixed in the rules form. These rules exclude such mistakes further repetition.
	
	The SRE best practices makes system more stable and reduces the production system outages.\newline
	mainly SRE should focus on below information to provide best support to customers, and make systems more stable with better performance,
	\begin{itemize}
		\item SLIs(Service Level Indicators)
		\item  SLOs(Service Level Objectives)
		\item  SLAs(Service Level Agreements)
		\item Application Monitoring
		\item Alerting mechanism when application is not functioning as expectecd. 
	\end{itemize}   
	
	\section*{Methodology}
	The primary research method for this study is literature review and SRE best practices along with useful case studies to make tasks easier to analyze and take action proactively before going the situation worse on the environment. Upon receiving the alert with respect to any production application as a first step logs have to be analyzed and identify issue is related to CPU or Disk or Memory, and most of the production applications will interact with database. So DB has to be monitored with respect to IOPS, read latency, write latency, CPU, LAG, db connection limit. \newline
	
	As a second step, when you receive a problem report or incident from the production environment, the next step is to figure out what has to be analyzed to find out actual Root Cause of the issue. Problems can vary in severity, an issue might affect only one user or multiple users under very specific circumstances and might have a workaround or not, or it might entail a complete global outage for a service. Your response should be appropriate for the problem’s impact, it’s appropriate to declare an all-hands-on-deck emergency for the latter but doing so for the former is overkill. Assessing an issue’s severity requires an exercise of good engineering judgment and, often, a degree of calm under pressure. After finding the RCA(Root Cause Analysis) of particular issue then share the RCA with customers and store it in internal bug tracing system and it is needed to apply the fix when the similar kind of issues are encountered in the future.\newline
	
	There are many ways to simplify and speed troubleshooting. Perhaps the most fundamentals are:
	\begin{itemize}
	\item Building the observability with both white-box metrics and structured logs of each component of application.
	\item Designing systems with well-understood and observable interfaces between components of the application.		
	\end{itemize}

    \section*{References}
	\begin{enumerate}
		\item Tariq Hussain Sheakh.2Dr S.M.K.Quadri, 3VijayPal Singh, A Critical Review of Software Reliability, International Journal of Emerging Technology and Advanced Engineering Website: www.ijetae.com (ISSN 2250-2459, Volume 2, Issue 4, April 2012)
		\item Jinho Hwang, Larisa Shwartz, Qing Wang, Raghav Batta, Harshit Kumar, Michael Nidd FIXME: Enhance Software Reliability with Hybrid Approaches in Cloud
		\item Irene Eusgeld, Falk Fraikin, Matthias Rohr, Felix Salfner, and Ute Wappler, 2005 Software Reliability
		\item Michael Carr AT\&T Advanced Software Construction Center Cary, NC 2751 1 USA A Data Analysis and Representation Engine to Support Software Reliability Engineering
	\end{enumerate}
    \begin{enumerate}
    \setcounter{enumi}{4} 

    \item Site Reliability Engineering: How Google Runs Production Systems.  
    Available at: \url{https://books.google.com/books?hl=en&lr=&id=_4rPCwAAQBAJ&oi=fnd&pg=PP1&ots=pAiM9yNXXm&sig=ob5T6514E_GkE92gG13c4wv6X1c#v=onepage&q&f=false}

    \item Google. SRE Book. Available at: \url{https://landing.google.com/sre/sre-book/chapters/effective-troubleshooting/}, 2020. [online].

\end{enumerate}
\end{document}